\documentclass[twocolumn,showpacs,superscriptaddress]{revtex4-1}

\usepackage{dcolumn}
\usepackage{bm}         
\usepackage{amssymb}
\usepackage{graphicx}

\begin{document}

\title{Direct current superconducting quantum interferometers with asymmetric shunt resistors}

\author{M.~Rudolph}
\author{J.~Nagel}
\affiliation{%
  Physikalisches Institut -- Experimentalphysik II and Center for Collective Quantum Phenomena in LISA$^+$,
  Universit\"at T\"ubingen,
  Auf der Morgenstelle 14,
  D-72076 T\"ubingen, 
  Germany
}
\author{J.M.~Meckbach}
\affiliation{%
  Institut f\"ur Mikro- und Nanoelektronische Systeme, 
  Karlsruhe Institute of Technology, 
  Hertzstr. 16, 
  D-76187 Karlsruhe, 
  Germany
}
\author{M.~Kemmler}
\affiliation{%
  Physikalisches Institut -- Experimentalphysik II and Center for Collective Quantum Phenomena in LISA$^+$,
  Universit\"at T\"ubingen,
  Auf der Morgenstelle 14,
  D-72076 T\"ubingen, 
  Germany
}
\author{M.~Siegel}
\affiliation{%
  Institut f\"ur Mikro- und Nanoelektronische Systeme, 
	Karlsruhe Institute of Technology, 
	Hertzstr. 16, 
	D-76187 Karlsruhe, 
	Germany
}
\author{K.~Ilin}
\affiliation{%
  Institut f\"ur Mikro- und Nanoelektronische Systeme, 
  Karlsruhe Institute of Technology, 
  Hertzstr. 16, 
  D-76187 Karlsruhe, 
  Germany
}
\author{D.~Koelle}
\author{R.~Kleiner}
\email{kleiner@uni-tuebingen.de}
\affiliation{%
  Physikalisches Institut -- Experimentalphysik II and Center for Collective Quantum Phenomena in LISA$^+$,
  Universit\"at T\"ubingen,
  Auf der Morgenstelle 14,
  D-72076 T\"ubingen, 
  Germany
}

\begin{abstract}
We have investigated asymmetrically shunted Nb/Al-AlO$_x$/Nb direct current (dc) superconducting quantum interference devices (SQUIDs). While keeping the total resistance $R$ identical to a comparable symmetric SQUID with $R^{-1} = R_1^{-1} + R_2^{-1}$, we shunted only one of the two Josephson junctions with $R = R_{1,2}/2$. Simulations predict that the optimum energy resolution $\epsilon$ and thus also the noise performance of such an asymmetric SQUID can be 3--4 times better than that of its symmetric counterpart. Experiments at a temperature of 4.2\,K yielded $\epsilon \approx 32\,\hbar$ for an asymmetric SQUID with an inductance of $22\,\rm{pH}$. For a comparable symmetric device $\epsilon = 110\,\hbar$ was achieved, confirming our simulation results. 
\end{abstract}


\maketitle
The transport characteristics and noise performance of direct current (dc) superconducting quantum interference devices SQUIDs having symmetric Josephson junctions has been intensively studied from the 1970's. 
Numerical simulations of the Langevin equations describing the SQUID dynamics reliably helped to understand the modulation patterns $V(\Phi_a,I)$ and the low-frequency voltage noise power $S_V(\Phi_a,I)$, where $V$ is the dc voltage across the SQUID, $I$ is the bias current and $\Phi_a$ is the applied flux. 
With the flux-to-voltage transfer function $V_\Phi  = |dV/d\Phi_a|$, one obtains the flux noise power $S_\Phi = S_V/V_\Phi^2$ or energy resolution $\epsilon = S_\Phi/2L$, where $L$ is the SQUID inductance. 
For an optimized device one obtains in the limit of small thermal fluctuations an energy resolution $\epsilon = (8-9) k_BTL/R$ for an inductance parameter $\beta_L = 2I_0L/\Phi_0$ somewhat below 1 \cite{Tesche77,Bruines82}. 
Here, $I_0$ and $R$ respectively denote the junction critical current and resistance. $\Phi_0$ is the flux quantum.
Although $\epsilon$ can be very low -- for example, in Ref.~\onlinecite{vanHarlingen82} a value of $\sim$ 3 $\hbar$ has been reported at 4.2\,K for a 2\,pH device -- one may ask whether or not it still can be improved by introducing asymmetries in the junction parameters or perhaps by adding new elements to the SQUID.
Early simulations have shown that asymmetries in the junction critical currents and resistances can enhance $V_\Phi$, although for the prize of asymmetric $V(\Phi_a)$  patterns \cite{Tesche77}. 
It has also been predicted that an additional damping resistor can enhance $V_\Phi$ \cite{Enpuku85a, Enpuku85b} . 
Several works addressed junction asymmetries and additional damping resistors in more detail \cite{Kleiner96,Mueller01,Testa01a,Testa01b, Testa01c,Kahlmann01}, with the result that the transfer function can be increased and flux noise be decreased. 
The above investigations, however, explored only a very limited range of parameters and often addressed devices where the symmetric counterpart was far from optimum.  

Let us start with a theoretical analysis, using the standard Langevin equations \cite{Tesche77} where the Josephson junctions are described by the resistively and capacitively shunted junction model\cite{Stewart68, McCumber68}. 
With $i = I/I_0$ the normalized currents through the junctions $k=1,2$ are given by
\begin{equation}
\frac{i}{2} \pm j = \beta_{c} (1\pm \alpha_c)\ddot{\delta_k} + (1\pm \alpha_r)\dot{\delta_k} + (1\pm \alpha_i)(\Phi)\sin(\delta_k)+i_{N,k}
\label{Eq:RSJ1}
\end{equation}
, $\alpha_c$, $\alpha_r$ and $\alpha_i$ denote the asymmetries in capacitance, resistance and critical current respectively. 
The junction critical currents are  $I_{0,k} = I_0(1\pm \alpha_i)$, their resistances $R/(1\pm \alpha_r)$ and their capacitances $C_k=C(1\pm \alpha_c)$. `$\pm$' refers to junctions 1 and 2, respectively. 
$\delta_k$ denotes the phase of junction $k$, $j = J/I_0$ is the normalized circulating current in the SQUID loop and  
$\beta_c = 2\pi I_0R^2C/\Phi_0$ is the Stewart-McCumber parameter. 
Dots denote derivative with respect to normalized time $\tau = \Phi_0/2\pi I_0R$. 
The normalized noise current $i_{N,k}$ has a spectral power density $4\Gamma$, with $\Gamma = 2\pi k_BT/I_0\Phi_0$. 
The $\delta_k$ are related by
\begin{equation}
\delta_2 - \delta_1 = 2\pi\Phi/\Phi_0 + \pi\beta_L (j + \frac{\alpha_L}{2}i)
\label{Eq:RSJ2}
\end{equation}
where $\Phi$ is the total flux through the SQUID. 
$L = L_1 + L_2$, where $L_1$ and $L_2$ are the inductances of the two SQUID arms, related to the inductance asymmetry $\alpha_L$ via $L_k = L(1\pm \alpha_L)/2$.  

From Eqs.~(\ref{Eq:RSJ1}) and (\ref{Eq:RSJ2}) one obtains the normalized dc voltage $v = V/I_0R$, and thus the current voltage characteristic (IVC) by taking the time average of $u=(\dot{\delta_1}+\dot{\delta_2})/2$. 
From a Fourier transform of $u$ one obtains the normalized correlation functions $s_v=S_V2\pi I_0R/\Phi_0^3$, $s_\phi=S_\Phi I_0R/(2\Phi_0k_BT)$ and $e=s_\phi/2\Gamma\beta_L$. 

The quantity we are interested in most is the optimized normalized energy resolution $e_{\rm{opt}}$, where optimization is done for some or even all SQUID parameters. 
Recently, we have performed a systematic optimization of the noise performance of the rf SQUID, optimizing all of its parameters \cite{Kleiner07a, Kleiner07b}. 
We now apply the same procedure to fully optimize $e$ of the dc SQUID, with respect to $i,\phi_a=\Phi_a/\Phi_0, \beta_L, \beta_c, \alpha_i, \alpha_c$ and $\alpha_r$, i.e., for a given value of one or some of these parameters we find all others so that $e_{\rm{opt}}$ is minimized. 
The inductance asymmetry $\alpha_L$ does not appear in the above list, since, for a given bias current, it only causes a phase shift in $v(\phi_a)$  and also the noise correlation functions. 
Since in practice the junction capacitance is always nonzero, the McCumber parameter  $\beta_c$ should be as large as possible to obtain large values of $I_0R$. 
It turns out that $\beta_c $ values below 0.8 are uncritical in the sense that the other parameters can be tuned so that $e_{\rm{opt}}$ attains its minimum value irrespective of $\beta_c $. 
%
%
%

%
\begin{figure}[h]
\includegraphics[width=8cm]{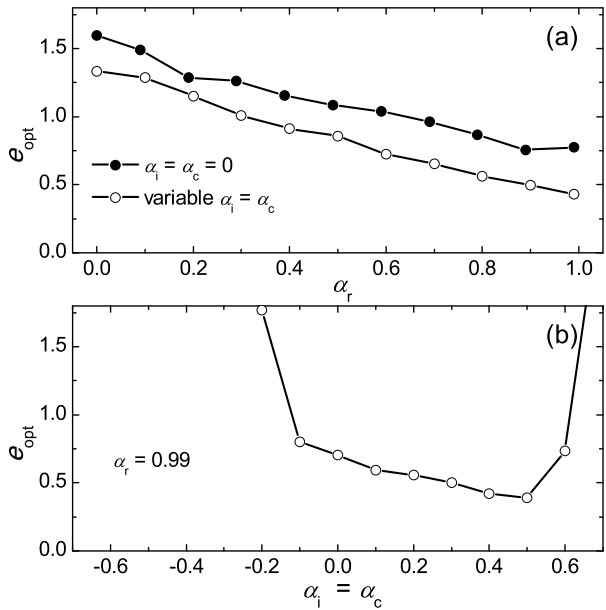}
\caption{Optimized normalized energy resolution $e_{\rm{opt}}$ vs. (a) resistance asymmetry $\alpha_r$ and (b) junction critical current and capacitance asymmetry $\alpha_i = \alpha_c$. Fixed parameters are $\Gamma = 0.01$ and $\beta_c = 0.7$. For all data points $\beta_L$ has been varied.}
\label{fig: Fig1}
\end{figure}
%
Fig.~\ref{fig: Fig1}(a) shows $e_{\rm{opt}}$ vs. $\alpha_r$ for $\alpha_i = \alpha_c$ = 0 (full circles) and for variable $\alpha_i = \alpha_c$ (open circles). 
Fixed parameters are $\beta_c$ = 0.7 and $\Gamma$ = 0.01. 
The parameters $i$, $\phi_a$ and $\beta_L$ have been varied to minimize $e$. 
We have used $\alpha_i = \alpha_c$, having in mind junctions where both $I_0$ and $C$ scale with the junction area and $R$ can be chosen independently by shunting. 
The value for $\Gamma$ was chosen as typical for operation at 4.2\,K. 
For $\alpha_i$ = 0, $e_{\rm{opt}}$ vs. $\alpha_r$  decreases from $\sim$ 1.6 to 0.7 for $\alpha_r \rightarrow $ 1. 
In case of variable $\alpha_i$ the minimum $e_{\rm{opt}}$ is about 0.4, i.e., a factor of 4 lower than the energy resolution of a comparable symmetric SQUID. 
Note that each point in the graph corresponds to different values of $i$, $\phi_a$, $\beta_L$.
We do not list the value of these parameters explicitly but note that in all cases $\beta_L$ was in the range $0.4-0.5$. $\alpha_i$ = $\alpha_c \approx0.3-0.5$ was found in case these parameters were varied.   

In Fig.~\ref{fig: Fig1}(b) we show  $e_{\rm{opt}}$ vs. $\alpha_i = \alpha_c$ for variable $\beta_L$ and fixed $\beta_c$ = 0.7 and $\Gamma$ = 0.01 and $\alpha_r = 0.99$. 
The lowest values of $e_{\rm{opt}}$ are achieved for $\alpha_i$ near 0.5. 
For lower values of $\alpha_i$, $e_{\rm{opt}}$ monotonically increases. 
In particular $e_{\rm{opt}}$ is achieved when $\alpha_i$ and $\alpha_r$ have the same sign, i.e., the junction having the \textit{lower} resistance should have the \textit{higher} $I_0$.  
We have obtained similar results, giving almost the same lowest values for $e_{\rm{opt}}$,  also for higher values of $\Gamma$ (up to 0.1). $e_{\rm{opt}}$ is thus a robust quantity.

In dimensioned units $\epsilon = e\cdot 2\Phi_0k_BT/I_0R$. To maximize $I_0$ for $\beta_L \approx 0.5$, $L$ should be as small as possible. 
Then, to maximize $R$ and keep $\beta_c$ below 1, $C$ should be as small as possible, which for a given capacitance per area means to keep the junction area as small. 
If junction asymmetries are considered, given a constant  critical current density, the size of the weaker junction is presumably limited by the fabrication process and, to obtain an $\alpha_i$ of, e.g., 0.3, the average junction area is increased by about 40$\%$ from its minimum value. 
This basically compensates the gain in $e_{\rm{opt}}$.
Asymmetries in $\alpha_i$ are thus not necessarily helpful. 
Thus, below we discuss an experimental design having $\alpha_i = \alpha_c = 0$.   

We also note that for $\alpha_r$ very close to 1, $e_{\rm{opt}}$ increases slightly again with increasing $\alpha_r$ for $\alpha_i = \alpha_c = 0$. 
This is related to chaotic dynamics which appears in some ranges of $i$ and $\phi_a$.  
Below, we will address this issue in comparison to experimental data. 
Nonetheless, if small values of $e_{\rm{opt}}$ can be retained for $\alpha_r \rightarrow 1$, the easiest way to realize the corresponding SQUID experimentally is to ``move'' the shunt from junction 1 to junction 2, leaving junction 1 unshunted and junction 2 shunted with a resistance $R$/2. 
In the following we discuss the performance of such a Nb/Al-AlO$_x$/Nb SQUID and compare it to simulations, as well as to the performance of a corresponding symmetric SQUID.
%
 \begin{figure}[h!]
\includegraphics[width=8cm]{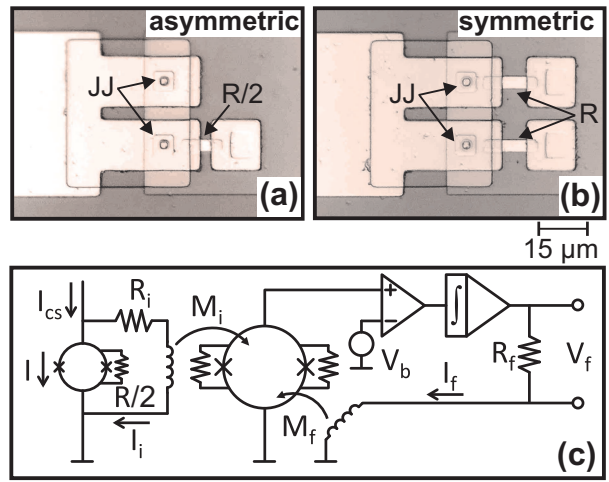}
\caption{(Color online) Optical image of the (a) asymmetric and (b) symmetric SQUID. The Josephson junctions are labeled by ``JJ'' and the shunt resistors by R and R/2.
(c) Readout scheme of the asymmetric SQUID using a SQUID amplifier operated in flux-locked loop.} 
\label{fig: Fig2}
\end{figure}
%
%
%

Figs.~\ref{fig: Fig2}(a) and (b) show optical images of an asymmetric and a symmetric SQUID. 
The SQUIDs have been fabricated using a Nb/Al-AlO$_x$/Nb technology based on optical photolithography. 
By a combination of reactive ion etching (RIE) employing CF$_4$ and O$_2$ and ion beam etching (IBE) the ground-electrode was defined. 
The subsequent definition of the junction area was done by RIE and anodic oxidation in an aqueous solution of (NH$_4$)B$_5$O$_8$ and C$_2$H$_6$O$_2$. 
Before the definition of the vias (RIE - IBE) the connecting bridges for the anodization between the individual SQUIDs were removed. 
The following definitions of the resistor, insulation layer and wiring layer were all done using a lift-off technique.
For the resistor material a $76\,\rm{nm}$ thick Palladium layer was deposited, resulting in a sheet resistance of $1\,\Omega$/sq at $T = 4.2 K$. The $\sim300\,\rm{nm}$ thick SiO insulation layer was deposited using thermal evaporation while the samples were mounted on a water cooled copper plate ($T_{\rm{process}} \le 26\,^\circ$C). 
After in-situ pre-cleaning, the final Nb wiring layer connecting the junctions, vias and shunt-resistors was dc-magnetron sputtered at room temperature.

Transport and noise measurements were performed at $T$ = 4.2\,K in a magnetically and electrically shielded environment. 
Dc characteristics (IVC, $V(\Phi_a)$, critical current $I_c(\Phi_a)$) were measured in a standard four-point configuration, using low noise current sources and a high impedance room temperature voltage amplifier (RTA).
For the noise measurements the RTA was not sensitive enough. 
Thus, $V$ was preamplified with a commercial SQUID amplifier \cite{Starcryo} having a $60\,\rm{pV}/\rm{Hz}^{1/2}$ resolution, operated in a flux-locked loop with ac flux bias at modulation frequency $f_{\rm{mod}}=256\,\rm{kHz}$.
The SQUIDs were operated open loop at fixed $I$ and $\Phi_a$. 
$V$ was measured by connecting the input coil of the SQUID amplifier in parallel to the SQUID. 
A $5\,\Omega$ resistor $R_i$ was in series to the input coil, as shown in Fig.\,\ref{fig: Fig2}(c). 
Due to the low input impedance of the amplifier, the current $I_{\rm{cs}}$ from the current source  divides into $I$ and the current $I_{\rm{i}}$ through the input coil. 
At given $I_{\rm{cs}}$, $I$ varies when changing $\Phi_a$, affecting $V_\Phi$ and thus the determination of the energy resolution. 
Using Kirchhoff`s laws and the condition for flux-locked loop operation, $I_i\rm{M}_i=\it{I}_f\rm{M}_f$, with $\rm{M}_i$  ($\rm{M}_f$) being the mutual inductance between the amplifier SQUID and the input (feedback) coil, one obtains $I_i=(\rm{M}_f/\rm{M}_i)(\it{V}_f/R_f)$. 
Since $\rm{M}_i$, $\rm{M}_f$ and $R_f$ are constants, $I_i$ can be determined by measuring $V_f$.
To determine $V(\Phi_a)$ for constant $I$ a software control-loop was implemented adjusting $I_{\rm{cs}}$ such that, for each value of $\Phi_a$, $I=I_{\rm{cs}}-\it{I}_i$ was fixed. 
The ratio $(\rm{M}_i/\rm{M}_f)$ slightly differed for different samples and was adjusted for each device until the bias corrected $V(\Phi_a)$ curve measured with the SQUID amplifier fitted the corresponding $V(\Phi_a)$ curve measured with the RTA; all other measurements for a given device were then performed with fixed $(\rm{M}_f/\rm{M}_i)$.
%
%
%
 
%
\begin{figure}[h]
\includegraphics[width=8cm]{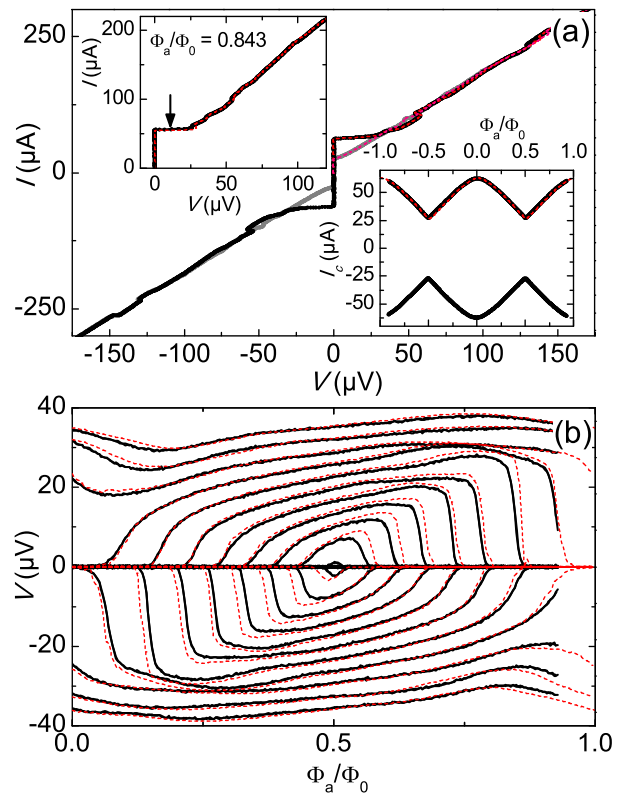}
\caption{(Color online) Dc characteristics of the asymmetric SQUID: (a) IVCs at $\Phi_a$ = 0 (solid black line) and at $\Phi_a = 0.5\,\Phi_0$ (solid gray line). Lower right inset: $I_c(\Phi_a)$. Upper left inset: IVC at $\Phi_a=0.843\,\Phi_0$. The arrow indicates the voltage for the lowest energy resolution. Theoretical curves are shown by dashed lines.
(b) $V(\Phi_a)$ (solid black line), for $I=-76.4\,\mu\rm{A} \ldots 75.9\,\mu\rm{A}$ (in $4.9\,\mu\rm{A}$ steps). Corresponding theoretical curves are shown by dashed lines.
}
\label{fig: Fig3}
\end{figure} 
%
%
%
Fig.~\ref{fig: Fig3} (a) shows IVCs of the asymmetric SQUID. 
Solid black line is for $\Phi_a=0$, solid gray line for $\Phi_a=0.5\,\Phi_0$. The critical current is $I_c \approx 2I_0 =I_{01}+I_{02}=62.0\,\mu$A and for $R/2$ we obtain 0.57\,$\Omega$, yielding $I_cR=35.3\,\mu$V.
One notes that, in contrast to IVCs of symmetric devices, the IVC of the asymmetric SQUID exhibits several structures, including regions of negative differential resistance. 
These structures are  reproduced in simulations, cf. dashed lines. 
The negative differential resistance in fact separates a high-current regime having chaotic dynamics from a more stable low-current regime.
For the simulation we have used parameters $\beta_L=0.675$, $\beta_c$ = 0.27, $\Gamma=0.0065$, $\alpha_r=0.999$, $\alpha_i=\alpha_c=0$. 
These parameters have been inferred partly by fitting the IVC, but also by fitting $I_c(\Phi_a)$. 
The corresponding data for $I_c(\Phi_a)$ are shown by solid black lines in the lower right inset of Fig.~\ref{fig: Fig3} (a). 
The dashed line in this graph shows the calculated curve. 
Finally, from $\beta_L$ and $I_0$ we obtain $L=21.7\,$pH which is close to the design value of $23.9\,$pH. 
The upper left inset of Fig.~\ref{fig: Fig3} (a) shows by solid black line the IVC taken at $\Phi_a=0.843\,\Phi_0$. 
For this particular flux value the best energy resolution was found at the voltage indicated by the arrow. 
The dashed line is a calculated curve, using the parameters given above.
A family of curves $V(\Phi_a)$ for variable $I$ is shown in Fig.~\ref{fig: Fig3}(b). 
Experimental data, for different $I$ are shown by solid black lines. 
The corresponding calculated curves, shown by dashed lines, fit the data reasonably well, showing that the dc characteristics of our device can be understood by the SQUID Langevin equations. 
In the $V(\Phi_a)$ curves one notes that the slope d$V$/d$\Phi_a$ is very steep for $\Phi_a\ge0.5\,\Phi_0$, in fact reaching maximum values of about $1.2\,\rm{mV}/\Phi_0$ near $I = 56\,\mu$A.
%
%
%

%
\begin{figure}[h]
\includegraphics[width=8cm]{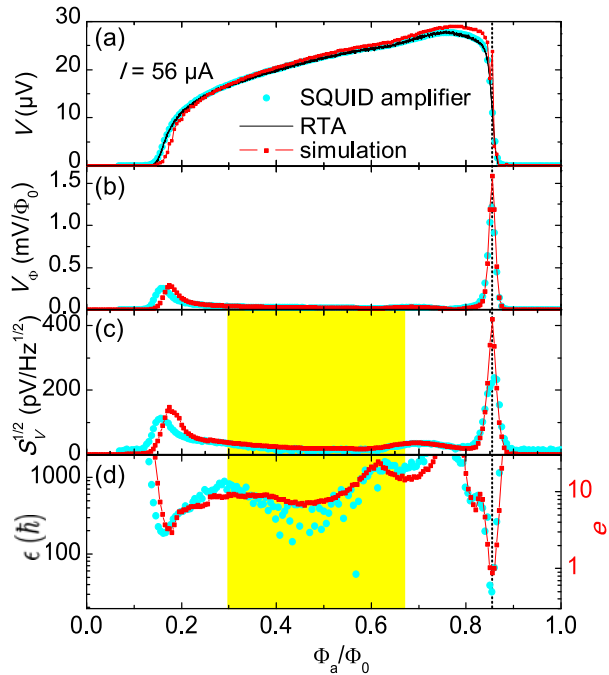}
\caption{(Color online) Electric transport and noise vs. $\Phi_a$ for the asymmetric SQUID at optimum $I =$56\,$\mu$A, measured with SQUID amplifier (dots) in comparison with numerical simulations (line plus symbol):
(a) Voltage across the SQUID; solid black line shows corresponding curve measured with a high impedance room temperature amplifier. 
(b) Transfer function $V_\Phi = $d$V$/d$\Phi_a$.
(c) Voltage noise $S_V^{1/2}$, experimental data averaged between $100\,\rm{Hz}\le \it{f} \le \rm{3}\,\rm{kHz}$ (white noise regime). (d) Absolute (left) and normalized (right) energy resolution. In (c) and (d) the noise of the SQUID amplifier has been subtracted; within the shaded area the SQUID amplifier noise was above the noise of the asymmetric SQUID, resulting in large errors when calculating $\epsilon$ and $e$. The vertical dotted line indicates the position of minimum $\epsilon$.
}
\label{fig: Fig4}
\end{figure} 
%
%
%
Our central results are shown in Fig.~\ref{fig: Fig4}. 
Figure \ref{fig: Fig4}(a) shows $V(\Phi_a)$ for the optimum bias current of 56\,$\mu$A.
Dots represent the experimental data, as taken by the SQUID amplifier. 
For comparison the solid black line represents the corresponding data taken by the RTA. 
The two curves are well on top of each other, justifying our method of correcting the bias current in measurements using the SQUID amplifier.
The line with symbols is the theoretical curve obtained from numerical simulations, which agrees well with the experimental curves. 
Figure \ref{fig: Fig4}(b) and (c), respectively, show by dots the experimental $V_\Phi$ and the white voltage noise $S_{V}^{1/2}$, and in comparison the corresponding calculated curves (lines plus symbols).  
For experimental data, $S_{V}^{1/2}$ has been averaged between $100\,\rm{Hz}\le \it{f} \le \rm{3}\,\rm{kHz}$. 
The voltage noise of the SQUID amplifier has been subtracted. 
Figure~\ref{fig: Fig4}(d) displays $\epsilon$ and $e$, calculated from the graphs shown in (b) and (c). 
For this device the optimum energy resolution is 32 $\hbar$; in normalized units, $e_{\rm{opt}} = 0.52$.
Surprisingly, the theoretical value for $e_{\rm{opt}}=0.85$ is higher. 
The reason for this is an instability in the calculations, appearing as a chaotic switching between two nearby voltage states \cite{Goldhirsch84}. 
This seems to be absent in the experimental device. 
The minimum rms flux noise was $133\,n\Phi_0/\rm{Hz}^{1/2}$ which is also quite low. 
For comparison, for the symmetric SQUID having parameters $I_0R=37.07\,\mu$V, $\beta_L=0.74$, $\beta_c=0.18$ and $\Gamma=0.00526$ we obtained $\epsilon_{\rm{opt}}=110\,\hbar$ ($e_{\rm{opt}}=1.79$) and $S_\Phi^{1/2}=361\,n\Phi_0/\rm{Hz}^{1/2}$, i.e., a factor of 3.4 higher $e_{\rm{opt}}$ than for our asymmetric device.
Further note that for the asymmetric SQUID $\beta_c$ was only 0.27, allowing in principle to increase $R$ by a factor of $\sim$ 1.5, potentially decreasing $\epsilon$ to $\sim$ 20 $\hbar$. 

In conclusion we have shown that asymmetries in the junction parameters of a dc SQUID can help to significantly improve its noise performance over a SQUID with symmetric junctions. 
The most practical way to achieve this is to leave one junction unshunted while the other junction is shunted by half of the resistance of the shunts of a symmetric device. 
Experimentally, we have found a factor of 3.4 improvement of the SQUID energy resolution over a symmetric device with comparable parameters. 
At least in simulations the drawback of the asymmetric shunt is to introduce chaotic behavior of the SQUID for certain regimes of bias current and applied flux. 
Our experimental device seems to be less sensitive to chaos. 
Thus, the asymmetrically shunted SQUID may be useful for applications where ultra-low values of energy resolutions are desired.

\acknowledgments
This work was supported by the European Research Council via ERC advanced grant SOCATHES, by the DFG via SFB/TRR
21 and in part by the DFG Centre of Functional Nanostructures Project N$^\circ$ B1.5. J.~Nagel acknowledges support by the Carl-Zeiss-Stiftung.

\end{document}